\documentstyle[prl,aps,multicol,epsf]{revtex}

\input{epsf}
\epsfverbosetrue
\epsfclipon

\newcommand{\putfigxsz}[4]{
   \begin{figure}\begin{center}\mbox{\epsfxsize #4
   \epsffile{#1}}
   \end{center}
   \end{figure}
   }

\begin{document}
\draft
\title{Metal-insulator transition at {\it B}=0
 in a dilute two dimensional GaAs-AlGaAs hole gas}
\author{M.Y. Simmons, A.R. Hamilton, M. Pepper, E.H. Linfield,
 P.D. Rose, and D.A. Ritchie}
\address{Cavendish Laboratory, Madingley Road, Cambridge CB3 OHE,
 U.K.}
\date{Originally submitted 22nd September 1997; updated 12 October 1997}
\maketitle
\begin{abstract}
We report the observation of a metal insulator transition at B=0 in a
high mobility two dimensional hole gas in a GaAs-AlGaAs
heterostructure. A clear critical point separates the insulating phase
from the metallic phase, demonstrating the existence of a well defined
minimum metallic conductivity $\sigma_{min}=2e^{2}/h$. 
The $\sigma(T)$ data either side of the transition can be `scaled' on
to one curve with a single parameter $T_0$. The application of a parallel
magnetic field increases $\sigma_{min}$ and broadens the
transition. 
We argue that strong
electron-electron interactions ($r_s \simeq 10$) destroy phase
coherence, removing quantum intereference corrections to the
conductivity.
\end{abstract}


\begin{multicols}{2}

In the mid-1970's experiments on Silicon inversion layers produced
considerable evidence for the existence of a metal-insulator
transition in 2D and a minimum metallic conductance,
$\sigma_{min}$ \cite{Pepper,Mott,Pepper2}. The decay constants of
localised state wavefunctions were investigated and it was shown that
when the number of localised electrons exceeded
$2{\times}10^{11}\text{cm}^{\text{-2}}$ the location of the mobility
edge was determined by electron-electron interactions and increased
with increasing carrier concentration. Subsequent theoretical work in
1979 suggested that all states in 2D were localised\cite{Abrahams} and
that phase incoherent scattering imposed a cut-off to a localised
wavefunction giving a logarithmic correction to metallic conduction
(weak localisation) which was widely observed and used to obtain very
detailed information on the various types of electron-electron
scattering in all three dimensions~\cite{Newson,Macfaden}. However in
order to investigate the
logarithmic correction at low, but accessible temperatures it was
necessary to use samples with low mobility so that the elastic
scattering length $l$ was small \cite{Uren}. In view of the success of 
the theory it was then assumed that the earlier high mobility samples 
did not show a logarithmic correction because the phase coherence length
$l_{\phi}$ was not greater than the elastic scattering length, but that 
if experiments could be performed at much lower temperatures (beyond the
capability of cryogenics) then the logarithmic correction would be
found.

Recent experimental results have raised this issue again and 
indicate that states in 2D are not always
localised with strong evidence for a metal-insulator transition in
high mobility Si MOSFETs \cite{Kravchenko}. It was found that the
resistivity on both the metallic and insulating sides of the
transition varied exponentially with decreasing temperature, 
and that a single scaling parameter could be used
to collapse the data on both sides of the transition onto a single
curve. Whilst the exact nature of the transition is
presently not understood there have been several reports of similar
scaling and duality between the resistivity and conductivity on
opposite sides of the transition, both for electrons in Si MOSFETs
\cite{Simonian,Popovic} and for holes in SiGe quantum wells
\cite{Coleridge}. In all of these reports electron-electron
interactions are known to be important, with the Coulomb interaction
energy being an order of magnitude larger than the Fermi energy at the
transition ($r_s \simeq 10$). The destruction of the metallic state by
an in-plane magnetic field has also lead to suggestions that spin
interactions are important \cite{simonian2,Pudalov}.

In this paper we present evidence of a metal-insulator transition at
B=0 in a high mobility, low density, two dimensional hole gas formed
in a GaAs-AlGaAs heterostructure.  The conductivity `scales' as a
function of temperature on both sides of the transition with a single
parameter $T_0$.  Normal metallic behaviour is observed for $\sigma >
\sigma_{min}$ in contrast to the exponential behaviour recently
observed in high mobility Si MOSFETs \cite{Kravchenko}. A parallel
magnetic field suppresses the metallic phase, demonstrating the
importance of spin interactions in this system.

The heterostructure used was grown by MBE
on a (311)$A$ GaAs substrate, and consisted of a 200~{\AA} GaAs
quantum well, modulation doped on one side with Si as the acceptor.
The carrier density $p_s$ was varied with an p$^+$
back-gate, formed using a combination of {\it in-situ}
ion-implantation and MBE regrowth \cite{Linfield}, 360~nm below the
quantum well.  Samples were processed into 450 by 50~$\mu$m Hall bars
aligned along the $[\bar2 33]$ direction and measurements were
performed in a $^3$He cryostat (with a base temperature of 270~mK)
designed for {\it in-situ} rotation of the sample with respect to the
magnetic field.  Low frequency (4~Hz) ac lockin techniques were used,
with excitations of 500~$\mu$V and 2~nA for two and four terminal
measurements respectively.  After illumination with a red LED the
carrier density could be varied in the range
$0-3.5{\times}10^{11}~\text{cm}^{\text{-2}}$ ($r_s>4$), with a peak
mobility of
$2.5{\times}10^{5}
~\text{cm}^{\text{2}}\text{V}^{\text{-1}}\text{s}^{\text{-1}}$,
over an order of magnitude larger than that used in previous studies.
For the carrier densities studied only the heavy hole subband is
occupied with $|M_J|$=3/2 (although for $k_{\parallel}\neq 0$ there is
some mixing between the light and heavy hole bands). The large effective 
mass ($m^*{\approx}0.3m_e$) quenches the kinetic energy thereby enhancing
the effects of Coulomb interactions. It should also be
noted that the asymmetric confining potential in our samples leads to
a partial lifting of the twofold Kramer's (spin) degeneracy away from
$k_{\parallel}=0$.

\begin{minipage}{7.2cm}
\noindent
\begin{figure}[tbph]
\putfigxsz{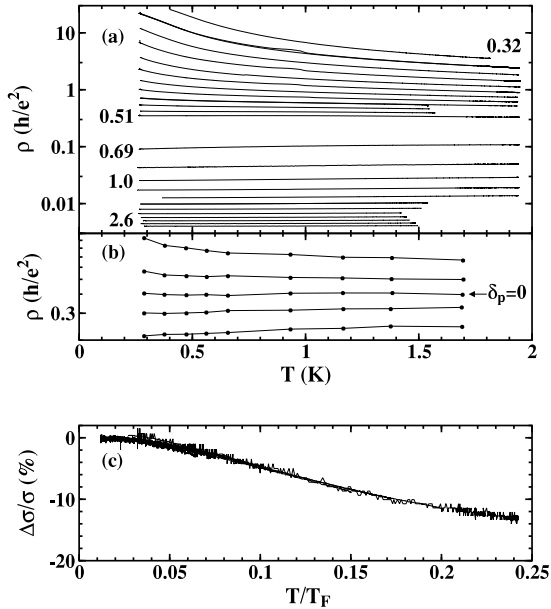}{lab}{cap}{7.2cm}
\caption{Temperature dependence of the resistivity: (a) as a function
of carrier density for
$p_{s}=0.32-2.6{\times}10^{11}~\text{cm}^{\text{-2}}$. (b) Close up of
the behaviour near the transition, showing the resistivity for 
$\delta_p=0, \pm5\%, \pm10\%$.  (c) Fractional change in conductivity 
$\Delta\sigma/\sigma$
against the normalised temperature $T/T_F$
for 11 equally spaced carrier densities in the range
$p_{s}=0.87-2.6{\times}10^{11}~\text{cm}^{\text{-2}}$.}
\label{fig1}
\end{figure}
\end{minipage}
\vspace{0.1in}

The transition from insulating to metallic behaviour with increasing
carrier density can be seen in the temperature dependence of the
resistivity in Fig. 1(a).  At a critical density
$p_{c}=5.1{\times}10^{10}~\text{cm}^{\text{-2}}$ ($r_s=11$) the
resistivity is temperature independent for $T{\lesssim}1.6K$ with
$\rho_{c} \simeq h/2e^{2}$, giving a minimum metallic conductance,
$\sigma_{min} \simeq 2e^2/h$.  At the lowest carrier densities
insulating behaviour is observed characterised by an exponential rise
in $\rho$ with decreasing temperature. In the strongly insulating
regime this behaviour fits $\rho(T)=\rho_{0}\exp{(T_{0}/T)^{1/2}}$
with $\rho_0 \simeq h/2e^2$, characteristic of variable range hopping
conduction in the presence of a Coulomb gap~\cite{Efros}.  Above this
critical density the resistivity changes behaviour to that of a normal
metal where $\partial\rho/\partial{T} \ge 0$ for all $T$.  Whilst
these curves demonstrate a clear metal-insulator transition, the large
decrease in resistivity observed for $T{\le}1.5K$ on the metallic side
of the transition in Si MOSFET and SiGe samples
\cite{Kravchenko,Coleridge} is not apparent.  Fig. 1(b)
shows the resistivity near the transition for carrier density changes
of $\pm5\%$ and $\pm10\%$ from $p_{c}$. Although no exponential
behaviour is observed on the metallic side for $T{\ge}0.3$~K and
$\delta_{p}{=}0.05$, it has been suggested that symmetry is only
expected to hold close to the transition where $\delta_p\equiv
(p_{s}-p_{c})/p_{c}\ll 1$ \cite{Dobrosavljevic}. We cannot therefore
exclude the possibility of exponential behaviour for $\delta_{p}\ll 1$
as $T\rightarrow 0$.

At high carrier densities, away from the transition, ($\delta_{p}\ge
0.2$) a gradual increase in the conductivity with temperature is
observed characteristic of normal metallic behaviour. In
Fig. 1(c) the fractional change in conductivity,
$\Delta\sigma/\sigma \equiv [\sigma(T)-\sigma(T{=}0)]/\sigma(T{=}0)$
is plotted against $T/T_F$ for carrier densities in the range
$0.87-2.6{\times}10^{11}~\text{cm}^{\text{-2}}$ \cite{sigma-T0}. We
find that $\Delta\sigma/\sigma$ scales as $T/T_F$, and is
approximately linear for $T/T_F{>}0.04$,
consistent with temperature dependent screening in the limit of low
disorder \cite{Gold}. It is noteworthy that a result derived for a
non-interacting system is applicable in a system where many body
interactions should be strong.  Close to the transition
($\delta_p{\leq}0.2$) this temperature dependence weakens and
$\Delta\sigma/\sigma$ deviates from the behaviour shown in
Fig.~\ref{fig1}(c), with $\sigma$ becoming completely temperature
independent at the critical point.

\begin{minipage}{7.2cm}
\noindent
\begin{figure}[tbph]
\putfigxsz{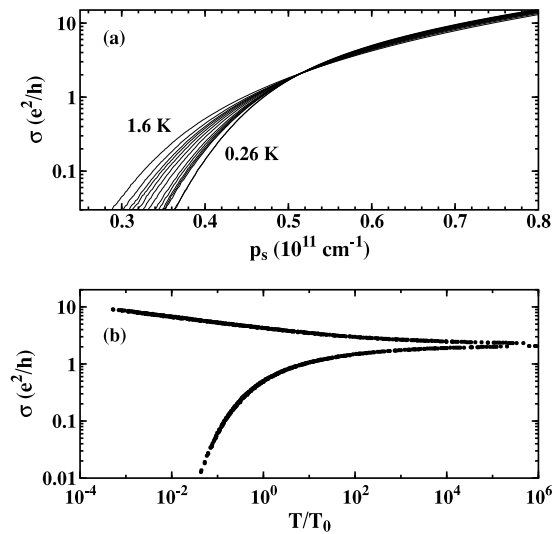}{lab}{cap}{7.2cm}
\caption{(a) Conductance as a function of carrier density for
different temperatures in the range 0.26-1.6~K, showing a clear
$\sigma_{min}=2e^2/h$ (a temperature independent contact resistance of
4.15~k$\Omega$ has been subtracted from all the data). (b) The scaled
data from (a) plotted against $T/T_{0}$. Hole densities are in the
range $0.35-0.68{\times}10^{11}~\text{cm}^{\text{-2}}$, $T<1.6$~K.}
\label{fig2}
\end{figure}
\end{minipage}
\vspace{0.1in}

Fig. 2(a) shows the conductivity as a function of gate
voltage at temperatures between 0.26 and 1.6~K. The curves all
intersect at a temperature independent point, confirming the
transition from insulating to metallic behaviour at a critical
conductivity of $\sigma_{min}{=}2e^{2}/h$.  The $\sigma(T)$ curves for
different carrier densities on the insulating side were made to
overlap by scaling them along the $T$ axis. For the lowest carrier
density ($p_s{=}0.35{\times}10^{11}~\text{cm}^{\text{-2}}$)
$T_{0}=6$~K was determined by fitting $\sigma(T) =
\sigma_{0}\exp{(T_{0}/T)^{-1/2}}$.  Each subsequent curve was then
individually scaled along the T-axis in order to collapse all the
curves onto a single trace, defining $T_0$ for each curve.  The
observation that data in the insulating regime can be scaled on to a
single curve is not surprising, since it is a direct consequence of
variable range hopping with a constant $\sigma_0$.
The same scaling
procedure was applied to the metallic data ($p_{s}\ge p_{c}$) where
the value of $T_{0}$ was chosen to be the same as that on the
insulating side close to the transition.  The results of this scaling
are presented in Fig. 2(b). We note that the scaling in the
metallic phase is less satisfactory than in the insulating phase (as
can also be observed in the data of
Refs.~\cite{Kravchenko,Kravchenko-Efield}), since at the higher
densities individual traces tend to flatten off at low temperatures,
as shown in Fig. 1(c).

\begin{minipage}{7.2cm}
\noindent
\begin{figure}[tbph]
\putfigxsz{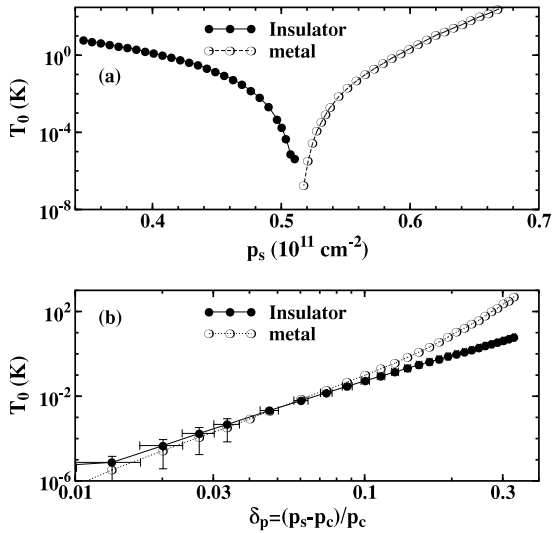}{lab}{cap}{7.2cm}
\caption{Scaling parameter $T_{0}$ (a) as a function of the hole
density, and (b) as a function of $\delta_p$.}
\label{fig3}
\end{figure}
\end{minipage}
\vspace{0.1in}

The scaling factor $T_0$ is shown in Fig. 3(a) as a function
of carrier density. For the lowest carrier densities 
$T_0$ is comparable to that observed in Si
MOSFETs \cite{Kravchenko}, but falls more rapidly as the transition is
approached.  Previous reports have found that the conductivity near 
the transition scales as 
\begin{equation}
\sigma(T,\delta_p) =  f(T/T_0) = f(|\delta_p|/T^{1/z\nu}),  
\label{eqnG1}
\end{equation} 
with a single parameter $T_{0}\propto |\delta_{p}|^{z\nu}$, where $z$
is the dynamical exponent and $\nu$ is the correlation length
exponent.  In our experiments the scaled data is not symmetric about
the critical $p_{s}$ and the second equality does not hold for 
$T/T_{0}\leq 1$. Fig. 3(b) shows $T_0$ against $|\delta_p|$.  
For $\delta_p{<}0.1$ the uncertainty in $T_0$ makes it difficult to
comment on the symmetry of $z\nu$ about the transition. At larger
$|\delta_p|$ the asymmetry is clearly visible, with $zv=3.8\pm0.4$ in
the insulating regime, and $7\pm1.5$ in the metallic regime.  In all
cases the values of $z\nu$ obtained are much larger than that observed
in Si-based samples where a universal value of $z\nu=1.6\pm0.2$ on
both the insulating and metallic sides of the transition has been
widely reported~\cite{Kravchenko,Simonian,Popovic,Coleridge}.
Physical insight into the variation of $T_{0}$ with $\delta_p$ is
obtained by considering the localisation length in the strongly
insulating regime, $\xi{=}e^{2}/\epsilon k_{B}T_{0}$ \cite{Efros}.
The localisation length therefore diverges as the transition is
approached. The large value of $z\nu$ shows that $\xi$ grows more
rapidly with increasing carrier density than in lower mobility
Si-based samples.  The reason for this difference is unclear, but may
be due to the long range of the random impurity potential in
modulation doped GaAs-AlGaAs heterostructures.

The scaling theory of localisation \cite{Abrahams} argues that there
is no $\sigma_{min}$ in the absence of spin-orbit scattering, as weak
localisation always takes over as $T{\rightarrow}0$. The introduction
of spin-orbit scattering leads to weak antilocalisation and the
possibility of a metal insulator transition. Although spin orbit
scattering is strong in p-GaAs we do not believe this to be the origin
of the metal insulator transition reported here. In our samples a
negative magnetoresistance is always observed in a perpendicular
magnetic field in contrast to the positive magnetoresistance expected
for antilocalisation. Despite the fact that we estimate
$l_{\phi}\approx 10~l$ no evidence of weak localisation or
antilocalisation is observed in the magnetoresistance near the
transition.  We suggest that the strength of the electron interactions
breaks phase coherence, removing the quantum interference corrections
to the conductivity.  The absence of these weak localisation
corrections thus restores the metal insulator
transition originally envisaged by Mott \cite{OldMott}, where
$\sigma(T{=}0)=0$ at $\sigma<\sigma_{min}$ and $\sigma(T{=}0)>0$ for
$\sigma\geq\sigma_{min}$, consistent with our data.

The application of a parallel magnetic field $B_{\parallel}$ couples
directly to the spin, altering many-body interactions and spin orbit
coupling by introducing a
spin-splitting of the `spin-up' and `spin-down' particles.  Although
the in-plane factor $g_{\parallel}$ is zero for purely heavy hole
states, mixing between the light and heavy hole bands at non-zero
$k_{\parallel}$ leads to a finite $g_{\parallel}$.  We have measured
the four terminal resistivity $\rho=1/\sigma$ as a function of
$B_{\parallel}$, and observe a negative magnetoconductance for all
carrier densities on both sides of the transition
(Fig.~\ref{fig4}(a)).  The effect of $B_{\parallel}$ on $\sigma_{min}$
is shown in Fig. 4(b), where we plot $\sigma(p_s)$ at
different temperatures for $B_{\parallel}=0$, 0.5, 1 and 3~$T$. The
critical point at which $\sigma=\sigma_{min}$ and is $T$-independent
can be seen to move to larger conductances as $B_{\parallel}$
increases, until at $B_{\parallel}=3$~T there is no distinct
transition between the metallic and insulating phases.  Increasing
$B_{\parallel}$ also makes the sample more insulating below the
transition ($\sigma$ is more $T$ dependent for a given $p_s$), and
weakens the metallic state on the other side of the transition
($\sigma(p_s)$ becomes less $T$ dependent).  In attempting to scale
the data according to Eqn.~\ref{eqnG1} we find that the scaling
exponents increase from $zv=3.75\pm0.25$ to $4.5\pm0.25$ in the
metallic regime, and from $6.5\pm1$ to $8\pm2$ in the insulating
regime, as $B_{\parallel}$ increases from 0 to 1~T, 
with a corresponding  decrease in the quality of
the scaling. The magnetic field has a 
dramatic effect on $\sigma_{min}$, indicated by the dashed 
lines in Fig. 4(b).  
As $B_{\parallel}$ increases $\sigma_{min}$ increases, and the
transition from an insulator to a metal is observed to broaden. In a
system with weak electron electron interactions and strong spin orbit
scattering the destruction of the metallic phase by the application of
a parallel magnetic field can occur as the spin degeneracy is lifted
and a transition from weak antilocalisation to weak localisation
occurs. However in our samples electron interactions should be strong
and no evidence for weak antilocalisation is
observed. The destruction of the metallic state
in a parallel field both in Si MOSFETs~\cite{simonian2,Pudalov} and in 
our samples does however point to a spin related origin of
the metallic phase.

\begin{minipage}{7.2cm}
\noindent
\begin{figure}[tbph]
\putfigxsz{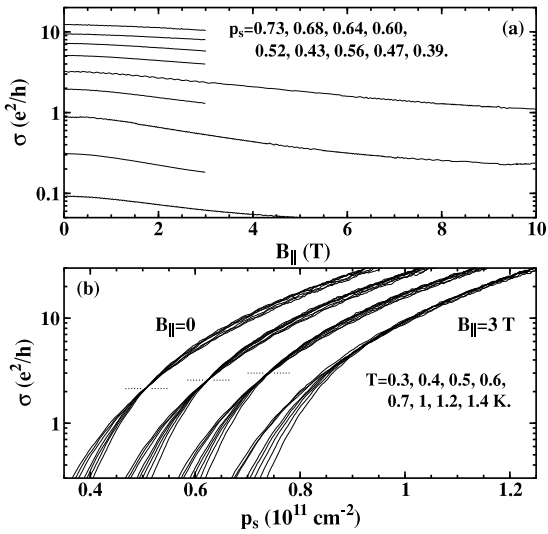}{lab}{cap}{7.2cm}
\caption{(a) Conductance as a function of applied parallel magnetic
field $B_{\parallel}$ at $T$=0.27~K for the hole densities indicated
on the graph (in unit of $10^{11}~\text{cm}^{\text{-2}}$).  (b) The
conductance as a function of carrier density, with magnetic fields of
$B_{\parallel}=$0, 0.5, 1 and 3~T. Curves for different
$B_{\parallel}$ have been offset horizontally by
$0.1{\times}10^{11}~\text{cm}^{\text{-2}}$; the horizontal dotted
lines mark the $T$-independent $\sigma_{min}$.}
\label{fig4}  
\end{figure}
\end{minipage}
\vspace{0.1in}

In summary we have reported the observation of a metal insulator
transition at B=0 in a high mobility two-dimensional hole gas formed
in a GaAs/AlGaAs heterostructure, with a minimum metallic conductance
of $\sigma_{min}=2e^2/h$.  Either side of the transition a single
scaling parameter can be used to collapse the resistivities onto a
single curve in both the conducting and insulating phases separately.
The critical exponents were found to be $7.0\pm1.5$ and $3.8\pm0.4$
respectively.  On the metallic side of the transition we observe
apparently normal metallic behaviour, with
$\Delta\sigma/\sigma=f(T/T_F)$. We suggest that this is a consequence
of the strength of the electron electron interactions $(r_s \approx
10)$ which remove phase coherent corrections to the conductivity.  
The spin related origins of the metallic state are however revealed by 
the application of a parallel magnetic field which suppresses the 
metallic phase and causes an increase in $\sigma_{min}$.

We acknowledge useful discussions with D. Popovi\'{c}, M. Kaveh,
R. Haydock, D.E. Khmel'nitskii, D. Neilson and C.J.B. Ford. This work
was funded by EPSRC (U.K.).


\end{multicols}

\begin{references}
\bibitem{sigma-T0}$\sigma(T{=}0)$ is
obtained by extrapolation to $T{=}0$, as $\sigma(T{<}0.5)$ is only a
weak function of $T$.
\end{references}
\end{document}